\documentclass[aps,showpacs,preprintnumbers,amsmath,amssymb]{revtex4}

\UseRawInputEncoding
\usepackage{graphicx}
\usepackage{epstopdf}
\usepackage{color}
\usepackage{subfig}
\usepackage{aasmacros}
\usepackage{accents}
\usepackage{natbib}

\begin{document}
    \captionsetup{justification=raggedright,singlelinecheck=false}

    \baselineskip=0.8cm
    \title{\bf Images of Kerr-MOG black holes surrounded by geometrically thick magnetized equilibrium tori}

    \author{Zelin Zhang$^{1}$\footnote{zzl@hunnu.edu.cn},
    Songbai Chen$^{1,2}$\footnote{Corresponding author: csb3752@hunnu.edu.cn},
    Jiliang Jing$^{1,2}$ \footnote{jljing@hunnu.edu.cn}}
    \affiliation{$^1$Department of Physics, Institute of Interdisciplinary Studies, Key Laboratory of Low Dimensional Quantum Structures
    and Quantum Control of Ministry of Education, Synergetic Innovation Center for Quantum Effects and Applications, Hunan
    Normal University,  Changsha, Hunan 410081, People's Republic of China
    \\
    $ ^2$Center for Gravitation and Cosmology, College of Physical Science and Technology, Yangzhou University, Yangzhou 225009, People's Republic of China}

    \begin{abstract}
    \baselineskip=0.6 cm
    \begin{center}
    {\bf Abstract}
    \end{center}
        We adopt general relativistic ray-tracing (GRRT) schemes to study images of Kerr-MOG black holes surrounded by geometrically thick magnetized equilibrium tori, which belong to steady-state solutions of thick accretion disks within the framework of general relativistic magnetohydrodynamics (GRMHD). The black hole possesses an extra dimensionless MOG parameter described its deviation from usual Kerr one.  Our results show that the presence of the MOG parameter leads to smaller disks in size, but enhances the total flux density and peak brightness in their images. Combining with observation data of black hole M87* from the Event Horizon Telescope (EHT), we make a constraint on parameters of the Kerr-MOG black hole and find that the presence of the MOG parameter broadens the allowable range of black hole spin.

    \end{abstract}

    \maketitle
    \newpage

    \section{Introduction}
     General Relativity (GR) has been a fundamental pillar of modern physics, essentially due to its perceived role in black hole physics and cosmology. It provides theoretical explanations for a lot of astronomical phenomena including Mercury's perihelion precession and the dynamics of binary pulsars \cite{2014LRR....17....4W}. With advancements in observational technologies, it is feasible to test GR under extreme conditions. Recent  black hole images releasing by the Event Horizon Telescope (EHT) collaborations \cite{2019ApJ...875L...1E} offer direct evidences of black hole existence in universe. Moreover, they also provide a powerful method to probe the electromagnetic interaction, matter distribution, and accretion process near black holes, and simultaneously examine validity of GR in strong gravitational fields \cite{Bardeen1973tla,Chandrasekhar1984siy,Cunha:2018acu,Perlick:2021aok,Liu:2022ruc,Wang:2022kvg,Vagnozzi:2022moj,Chen2023wzv,Chen2023wna,chen2022scf}. However, the current observations cannot completely exclude the possibility of the deviations from GR, which leaves an ample room for other alternative theories of gravity.

     The scalar-tensor-vector gravity (STVG) model \cite{2006JCAP...03..004M} is a kind of fully covariant modified gravity (MOG) theories. In this MOG theoretical model, there are three gravitational fields:  massless tensor graviton, a massless scalar graviton, and a massive vector graviton. It has successfully explained the rotation curves of galaxies  \cite{2013MNRAS.436.1439M, 2020MNRAS.496.3502D, 2024MNRAS.527.2687M} and the dynamics of galactic clusters \cite{2009MNRAS.397.1885M, 2014MNRAS.441.3724M, 2016arXiv160609128I}. The Kerr-like black hole solution in STVG theory is obtained, which owns an extra MOG parameter described the deviation from GR and yielded a variable gravitational constant \cite{2015EPJC...75..175M}. Moreover, the MOG parameter describes a new degree of freedom, which influences thermodynamic properties of the black hole \cite{2016PhLB..757..528M}, dynamical behaviors of particles near the black hole \cite{2017EPJC...77..363S,2017EPJC...77..655L,2018PhRvD..97l4049S,2020EPJC...80..399H,2021Galax...9...75R,2023Symm...15.2084M,2023arXiv231116936U}, and quasinormal modes \cite{2018PhLB..779..492M,2023JCAP...11..057L}. It is of interest to study the observational effects of the Kerr-MOG black hole because they could help understand this modified gravity theory. With observations of the S2 star orbiting the supermassive black hole at the Milky Way's center, the MOG parameter $\alpha$ is constrained to $\alpha \lesssim 0.410$ \cite{2022Univ....8..137D}. With the  black hole shadows \cite{2015EPJC...75..130M,2019JCAP...03..046W,2020PhRvD.101b4014M,2022EPJC...82..885H}, the allowed range of the upper limit of the MOG parameter $\alpha_{up}$ is given by $0.350\lesssim\alpha_{\rm up}\lesssim 0.485$ for the M87* black hole and $0.162 \lesssim \alpha_{\rm up} \lesssim 0.285$ for the SgrA* black hole \cite{2022PhRvD.106f4012K}.

    Accretion disks are actual and feasible light sources around black holes so the properties of accretion disks have a much greater impact on black hole images. Adopting thin accretion disk model around the Kerr-MOG black hole, it is found that the MOG parameter $\alpha$ influences radiation intensity and the appearance of the black hole images \cite{2017PhRvD..95j4047P,2023EPJC...83..264H}. Moreover, the effects of the MOG parameter $\alpha$ on polarized images arising from a thin accretion disk  has been investigated for the rotating black hole \cite{2022ApJ...938....2Q}. However, above literatures focus only on the thin accretion disk models around MOG black holes. Thus, how thick accretion disk models affect images of the Kerr-MOG black hole is still an open issue. It is well known that studying images of a black hole surrounded by a thick accretion disk must solve general relativistic magnetohydrodynamic equations of fluids around the black hole. In general, such kind of dynamical equations are very complex so it is very difficult to an analytical solution for them. Thus, the high precise simulation black hole images illuminated by a thick accretion disk must resort  numerical GRMHD codes, which expends a significant amount of computing resources. Therefore, semi-analytic models remain favored\cite{2003ApJ...598..301Y, 2018ApJ...863..148P, 2022A&A...667A.170V, 2024arXiv240504749C,2024JCAP...05..032Z,2024JCAP...02..030H,2011ApJ...735..110B,2024JCAP...01..059J}. Magnetized equilibrium tori is a kind of simple theoretical models of steady-state geometrically thick accretion disks, which can reveal basic configurations and main features of thick accretion disks through analytical methods. Thus, magnetized equilibrium tori have been applied to simulate millimeter images of Sgr A*, successfully reproducing the spectral constraints in the millimeter domain \cite{2015A&A...574A..48V}. The purely toroidal magnetic field, assumed as a fundamental assumption in magnetized equilibrium tori, can generate large-scale poloidal magnetic flux, which is necessary to power jets \cite{2020MNRAS.494.3656L}. Furthermore, the magnetized equilibrium tori model has also been studied in the spacetimes of deformed compact objects \cite{2021A&A...654A.100F}, boson stars \cite{2023PhRvD.107j3043G}, naked singularities \cite{2023CQGra..40s5011P}, dyonic black holes in quasi-topological electromagnetism \cite{2024EPJC...84..130Z}, and Schwarzschild black holes in swirling universes \cite{2024arXiv240202789C}. In this paper, the main motivation is to study the Kerr-MOG black hole images illuminate by such geometrically thick magnetized equilibrium tori and to probe effects of the MOG parameter on the images. Note that although Kerr-MOG black holes may share similarities with Kerr-Newman black holes, such images have not yet been studied.

    The paper is organized as follows: In Sec.II, we  present an analysis of the magnetized equilibrium tori surrounding a Kerr-MOG black hole. In Sec.III, we simulate images of a Kerr-MOG black hole with magnetized equilibrium tori and to probe effects of the MOG parameter on the images. Finally, we end the paper with a summary.

    \section{Magnetized Equilibrium Tori around a Kerr-MOG Black Hole  }
    \label{sec:2}

    The Kerr-MOG black hole is a stationary axisymmetric solution of field equations within the MOG framework of STVG \cite{2015EPJC...75..175M}, and its metric form in Boyer-Lindquist coordinates can be expressed as
    \begin{eqnarray}
        \label{metric}
        ds^2 = &-&\left(\frac{\Delta-a^2\sin^{2}\theta}{\Sigma}\right) dt^2+\frac{\Sigma}{\Delta}dr^{2}+\Sigma d\theta^2 -2\left(\frac{r^2+a^2-\Delta}{\Sigma}\right)a \sin^{2}\theta dt d\phi \\
        &+&\left[\frac{\left(r^2+a^2\right)^2-\Delta a^{2}\sin^2\theta }{\Sigma}\right]\sin^2\theta d\phi^2,
    \end{eqnarray}
    where
    \begin{eqnarray}
        \Delta&=&r^2-2GMr+a^2+\alpha G G_N M^2,\\
        \Sigma&=&r^2+a^2\cos^2\theta.
    \end{eqnarray}
    The parameters $a$ and $M$ represent the black hole's spin and mass, respectively. The MOG parameter $\alpha$ adjusting gravitational constant strength as $G = (1+\alpha) G_N$, where $G_N$ is Newton-gravitational constant. Form now onwards, we adopt $G_N = 1$. The Arnowitt-Deser-Misner (ADM) mass $M_\alpha $ is related to the mass parameter $M$ by $M_\alpha = (1+\alpha)M$ \cite{2018PhRvD..97l4049S}. With the ADM mass $M_\alpha$, the outer and inner horizon radii are roots of the equation
    \begin{equation}
        \Delta = r^2 - 2M_{\alpha} r + a^2 + \frac{\alpha}{(1+\alpha)}M_{\alpha}^2 = 0.
    \end{equation}
    and their forms are
    \begin{equation}
        r_{\pm} = M_{\alpha} \pm \sqrt{\frac{M_{\alpha}^2}{(1 + \alpha)} - a^2}.
    \end{equation}
    Moreover, the extremal limit for the Kerr-MOG black hole is $M_\alpha ^2 = (1+\alpha)a^2$.

    To study images of the Kerr-MOG black hole surrounded by geometrically thick magnetized equilibrium tori, we must firstly consider general relativistic magnetohydrodynamics in the curved spacetimes where fluid is dominated by three conservation equations as follows \cite{2006MNRAS.368..993K},
    \begin{eqnarray}
        \nabla_\mu T^{\mu \nu} = 0,\quad\quad\quad
        \nabla_\mu {}^*\!F^{\mu \nu} = 0,\quad\quad\quad
        \nabla_\mu (\rho u^\mu) = 0,\label{tcon1}
    \end{eqnarray}
    where $^*\!F^{\mu \nu}$ is the Faraday tensor and $\rho$ is the rest mass density of fluid.  $T^{\mu \nu}$ is the energy-momentum tensor with the form
    \begin{equation}
        T^{\mu \nu} = (h+b^2) u^\mu u^\nu + (p+\frac{1}{2} b^2) g^{\mu \nu} - b^\mu b^\nu, \label{tcon2}
    \end{equation}
    where $h$ and $p$ correspond to the fluid enthalpy and pressure, respectively. $u^\mu$ is the four velocity of fluid and $b^\mu$ is the 4-vector form of the magnetic field. Specifically, in the fluid frame, $b^\mu$ takes the form $(0, B)$, where $B$ is the usual three-vector of the magnetic field as measured in the fluid frame. For the sake of simplicity, here we assume that
    the fluid is a barotropic fluid with positive pressure and without self-gravity.  Moreover, the fluid is assumed to be axisymmetric and stationary, and the rotation of perfect fluid is restricted to be in the azimuthal direction and the magnetic field in the spacetime is also assumed to be purely azimuthal. With these assumptions, the four velocity of fluid and the four component of the magnetic field can be respectively expressed in the black holes frame as
   \begin{equation}
        u^{\mu}=(u^t,0,0,u^{\phi}),\quad\quad\quad\quad b^{\mu}=(b^t,0,0,b^{\phi}).\label{tcon3}
    \end{equation}
    Inserting Eq. (\ref{tcon3}) into Eqs.(\ref{tcon1}) and (\ref{tcon2}), one has  \cite{2006MNRAS.368..993K}
    \begin{equation}
        \partial_\nu\left(\ln |u_t|\right) - \frac{\Omega}{1-l \Omega} \partial_\nu l + \frac{\partial_\nu p}{h} + \frac{\partial_\nu\tilde{p}_m}{\tilde{h}} = 0,
        \label{eq:3-5}
    \end{equation}
    where $\tilde{h} = \mathcal{L} h$ and $\tilde{p}_m = \mathcal{L} p_m$.  The factor $\mathcal{L}$ is related to the black hole metric by $\mathcal{L} = g_{t\phi}g_{t\phi}-g_{tt}g_{\phi\phi}$. $p_m$ is the magnetic pressure with the form $p_m = b^2/2$. Here, $\Omega = u^{\phi}/u^{t}$ and $l = - u_{\phi}/u_t$ are respectively the angular velocity and specific angular momentum of the particles in fluid.
    As in \cite{2006MNRAS.368..993K}, we further assume that the fluid in the magnetized equilibrium tori presents a constant specific angular momentum $l = l_0$ and the corresponding  equations of state are
    \begin{equation}
        p=K h^\kappa, \quad \quad \quad \tilde{p}_m = K_m \tilde{h}^\eta,\label{eq:3-9}
    \end{equation}
    where $\kappa$ and $\eta$ are two constants.
    With these assumptions, equation (\ref{eq:3-5}) can be integrated as \cite{2006MNRAS.368..993K}
    \begin{equation}
       W-W_{in} + \frac{\kappa}{\kappa-1} \frac{p}{h} + \frac{\eta}{\eta-1} \frac{p_m}{h} = 0.
       \label{eq:3-11}
    \end{equation}
    Here $W$ is the potential with the form
    \begin{equation}
        W(r,\theta)=\frac{1}{2}\ln\left|\frac{{\cal L}}{{\cal A}} \right|.
    \end{equation}
    where ${\cal A} = g_{\phi\phi}+2l_0g_{t\phi}+ l_0^2 g_{tt}$. The quantity $W_{in}$ denotes the value of the potential at the inner edge of the disk. With these equations, we can analyze the structures and properties of geometrically thick magnetized equilibrium tori around the Kerr-MOG black hole.

   The value of $l_0$ plays a pivotal role in the potential $W$ which determines the topology and structures of the equilibrium tori.
   Here we set the constant specific angular momentum $l_{ms} < l_0 < l_{mb}$,  where $l_{ms}$ and $l_{mb}$ are respectively the specific angular momentum of the fluid particles moving along the marginally stable circular orbit $r_{ms}$ and the marginally bound orbit $r_{mb}$. The main reason is that the existence of magnetized equilibrium tori around black holes requires that the value of $l_0$ in the fluid must satisfy $l_0>l_{ms}$ \cite{2006MNRAS.368..993K}. Meanwhile, the condition $l_0 < l_{mb}$ ensures that the equilibrium tori have a cusp and a center, which could bring richer features for black hole images. Therefore,
   the potential $W(r,\theta)$ inside the disk usual satisfies $W_c < W(r,\theta) < W_{cusp}$, where $W_c$ and $W_{cusp}$ respectively correspond to the potential at the disk's center and at the cusp. For the convenience, one can introduce the dimensionless specific angular momentum
    \begin{equation}
        \lambda \equiv \frac{l_0-l_{ms}}{l_{mb}-l_{ms}},
    \end{equation}
    and the dimensionless potential
    \begin{equation}
        w(r,\theta) = \frac{W(r,\theta)-W_{cusp}}{W_c-W_{cusp}}.
    \end{equation}
    It is obvious that these two dimensionless parameters satisfy $0<\lambda<1$ and $0 \leq w(r,\theta) \leq 1$, respectively. Similarly, we recast $W_{in}$ in terms of the dimensionless parameter as $w_{in} = (W_{in}-W_{c})/(W_{cusp}-W_c)$. In order to obtain the distribution of enthalpy $h$ and pressure $p$ in the magnetized equilibrium tori for fixed $\kappa$ and $\eta$, we must firstly get the gas pressure  and the magnetic pressure at the disk's center $r=r_c$, i.e.,  \cite{2006MNRAS.368..993K}
    \begin{equation}
       p_c = h_c w_{in} (W_{cusp}-W_c)
      \left(\frac{\kappa}{\kappa-1} +
            \frac{\eta}{\eta-1} \frac{1}{\beta_c}
       \right)^{-1},\quad\quad\quad  p_{m_c} = p_c / \beta_c,
       \label{eq:pc}
    \end{equation}
  where $h_c$ and  $\beta_c = (p/p_m)_c$ denote  the enthalpy and the plasma magnetization parameter at the disk's center, respectively. With Eqs.(\ref{eq:pc}), the constants $K$ and $K_m$ in equation (\ref{eq:3-9}) can be given.  Solving Eq.(\ref{eq:3-11}), we can get the distribution of enthalpy $h$ and pressure $p$ in the magnetized equilibrium tori. The rest mass density in the tori can be further given by
    \begin{equation}
        \rho = h - \frac{\kappa p}{\kappa - 1}.
    \end{equation}
    \begin{figure}
        \centering
        \includegraphics[width=16cm]{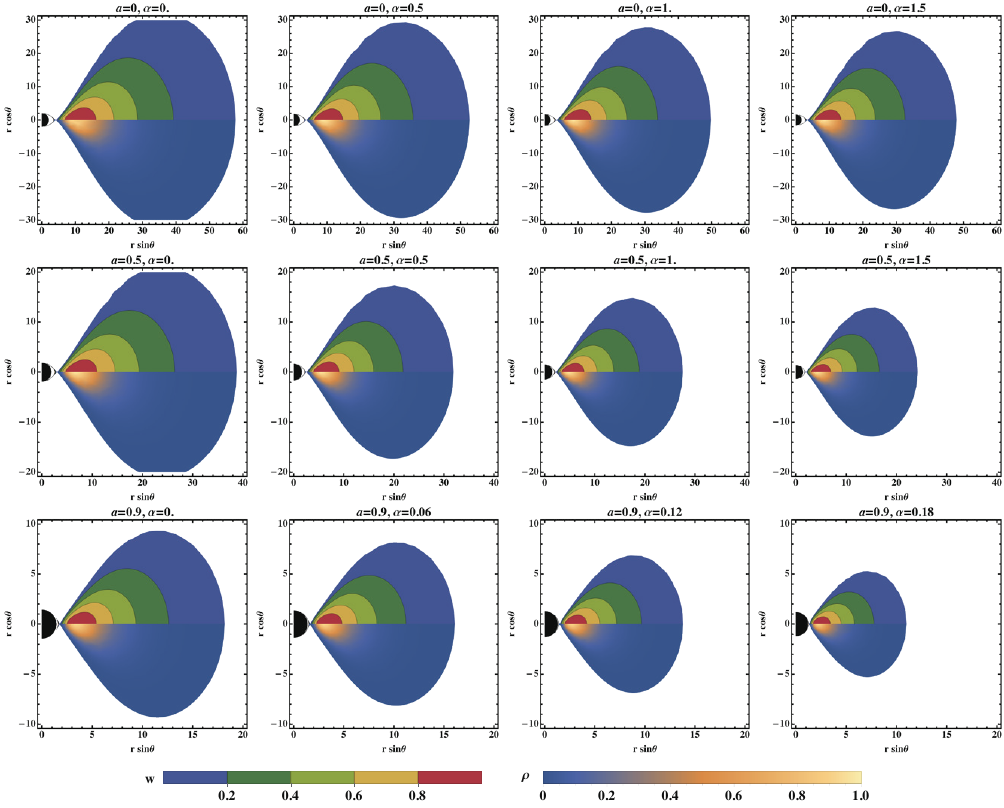}
        \caption{Dimensionless potential and density distribution in the geometrically thick magnetized equilibrium tori around Kerr-MOG black holes with different spin and MOG parameters. In each panel, the upper half and the lower one depict the dimensionless potential and the rest mass density distribution, respectively.}
        \label{fig:1}
    \end{figure}
   In general, the regions in accretion disk around black holes is  weakly magnetized accretion disks with high $\beta$, where $\beta \equiv p / p_m$ is the plasma magnetization parameter \cite{2016A&A...586A..38M, 2019ApJ...875L...5E}. Fig. \ref{fig:1} presents the geometrically thick magnetized equilibrium tori around the Kerr-MOG black hole, where we set $\beta_c=10$,  $h_c = 1$, $w_{in} = 1$, $\lambda = 0.8$ and $\kappa = \eta = 4/3$. One can find that the dimensionless potential $w$  reaches its maximums at the tori centers and drop to zero at its outermost surface. The rest mass density distribution is similar to that of the potential $w$, but the rest mass density $\rho$ own more rapidly rate of decay with the distance from the center. Moreover, both the black hole spin $a$ and the MOG parameter $\alpha$ lead to the sharp decrease in the size of equilibrium tori. However, the shape of equilibrium tori is not susceptible to these two parameters.

    \section{Simulated Images of Magnetized Equilibrium Tori}
    \label{sec:3}

   To simulate images of geometrically thick magnetized equilibrium tori around the Kerr-MOG black hole,  here we assume that the equilibrium tori are low-density and high-temperature,  where the plasma is collisionless and the electron temperature $T_e$ is lower than the ion temperature $T_i$ as in \cite{2014ARA&A..52..529Y}. The electron temperature directly affects synchrotron radiation emission in the tori.
   Following the model in \cite{2019ApJ...875L...5E}, the electron temperature $T_e$ is assumed as
    \begin{equation}
        T_e = \frac{2 m_p u}{3 k_B \rho (2+R)},
    \end{equation}
    which depends on the plasma density $\rho$, the internal energy density $u$, the magnetization parameter $\beta$ and the proton mass $m_p$.
    Here $k_B$ is Boltzmann's constant and
     $R \equiv T_i/T_e$ is the proton-to-electron temperature ratio with a form
    \begin{equation}
        R = R_{high} \frac{\beta^2}{1+ \beta^2} + \frac{1}{1+\beta^2},
    \end{equation}
    which means that the ratio $R$ can be adjusted by the parameter $ R_{high} $ and the plasma magnetization parameter $\beta$ in the tori. The internal energy density $u$ can be obtained by the pressure $p$ using the relation $u = p/(\kappa - 1)$. As in \cite{2019ApJ...875L...5E}, for matching simulation images with actual observed images, one can introduce a scaling factor $\mathcal{M}$, which can adjust physical properties in magnetized equilibrium tori including
    the rest mass density, the magnetic field strength and the internal energy density, and further impacts the flux density in the simulated images.

    To numerically simulate the radiative transfer within the magnetized equilibrium tori around black holes, we utilize the \texttt{ARCMANCER} library \cite{2018ApJ...863....8P}, where the emission, absorption, and Faraday mixing coefficients of radiation in the magnetized plasma are calculated by employing the \texttt{symphony} code \cite{2016ApJ...822...34P,2018ApJ...868...13P,2021ApJ...921...17M}.
    Here, we focus only on the case where electrons are in thermal distributions and the corresponding radiative flux is produced only by thermal synchrotron emission.
    In this work, we set $R_{high} = 10$ to model the electron temperature and
    \begin{figure}
        \centering
        \includegraphics[width=16cm]{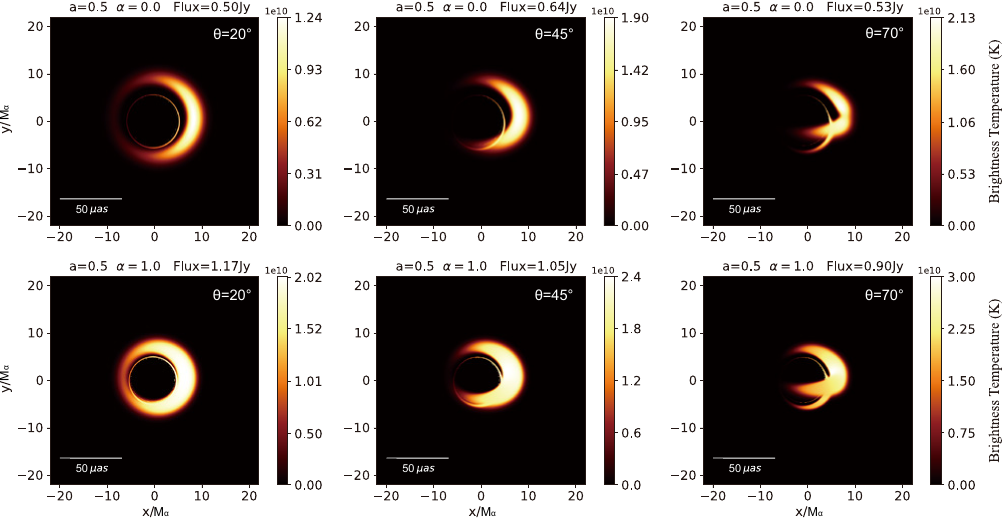}
        \caption{Simulated images of geometrically thick magnetized equilibrium tori around Kerr-MOG black hole for fixed spin parameter $a = 0.5$ with different MOG parameter $\alpha$ and observed inclination angle $\theta$.
       In each image,  the identical rescale parameter $\mathcal{M}$ is adopted to make comparisons. The images are presented in units of brightness temperature $T_b = S \lambda^2 / (2 k_B \Omega)$, where $S$ denotes the flux density, $\lambda$ is the observing wavelength and $\Omega$ the solid angle of the resolution element.}
        \label{fig:3}
    \end{figure}
     adopt the M87* black hole as a target with black hole mass $6.5 \times 10^9 M_{\odot}$ and the observer's distance $r_0 = 16.8 \  \mathrm{Mpc}$ \cite{2019ApJ...875L...1E, 2024A&A...681A..79E}. Moreover, in our simulations, the black hole images at the observation frequency $230 \ \mathrm{GHz}$ are set to be optically thin, which is consistent with the emission around low-luminosity active galactic nuclei (LLAGNs) emitted by the geometrically thick and optically thin disc.

 \begin{figure}
        \centering
        \includegraphics[width=6cm]{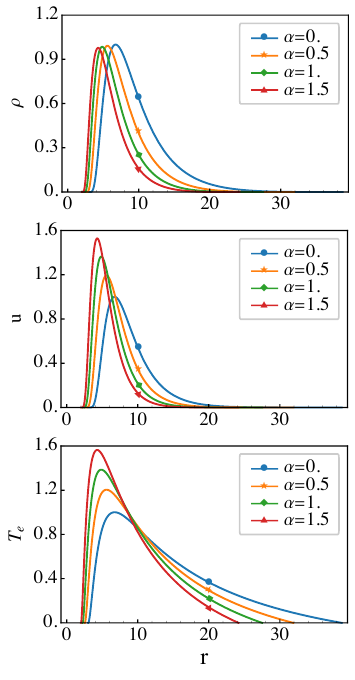}
        \caption{ Radial distributions of plasma density \(\rho\), internal energy density \(u\), and electron temperature \(T_e\) in the tori across equatorial plane of the black hole with the fixed spin \(a = 0.5\). The values of \(\rho\), \(u\), and \(T_e\) are normalized to their respective maximum values at \(\alpha = 0\). }
        \label{fig:2}
    \end{figure}
\begin{figure}
        \centering
        \includegraphics[width=8cm]{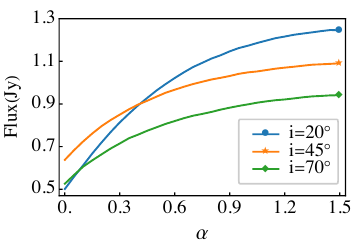}
        \caption{ Radiative flux varies with the \(\alpha\) parameter in the simulated images for \(a = 0.5\), using the same rescale parameters \(\mathcal{M}\) as in Fig. \ref{fig:3}.}
        \label{fig:4}
    \end{figure}

     Fig. \ref{fig:3} presents the simulated images of the Kerr-MOG black hole for fixed spin parameter $a=0.5$ with different MOG parameter $\alpha$ and observed inclination angle $\theta$. Here, in each image, the identical rescale parameter $\mathcal{M}$ is adopted to make comparisons.
     After setting the density scale,  we can numerically simulate the flux density $S$ for each  pixel and then the brightness temperature can be determined by $T_b = S \lambda^2 / (2 k_B \Omega)$, where $\lambda$ is the observing wavelength and $\Omega$ the solid angle of the resolution element.
     The images show that there are closely-packed bright thin rings together with a crescent-shaped bright zone. The crescent-shaped bright zone corresponds to the primary image of the magnetized equilibrium tori. The bright thin rings denote higher-order images of tori arising from light rays crossing the tori multiple times before reaching observation. Fig. \ref{fig:3} shows that the emission from the equilibrium tori dominates to make a ring-like emission, which is caused by that we here only focus on the emission from the equilibrium tori and
     do not consider those from the accretion flows near horizon continuously falling onto the black hole.
     From Fig. \ref{fig:3}, for different observer's inclination angle, we find that with the increasing of the MOG parameter $\alpha$, both the brightness temperatures and the radiative flux increase, but the size of the images decreases. Moreover, we also find that the width of the crescent-shaped bright zone increases with $\alpha$. The decreasing of the image size with the MOG parameter $\alpha$ is understandable because the sizes of equilibrium tori decrease with parameter $\alpha$. Generally, the radiative flux from the synchrotron radiation depends on the matter distribution and the electron temperature. In Fig. \ref{fig:2}, we present the changes of the plasma density $\rho$, internal energy density $u$, and electron temperature $T_{e}$ with the MOG parameter $\alpha$ in the equatorial plane of the black hole with the fixed spin $a=0.5$, which shows that the peak value of $\rho$ slightly decreases with the MOG parameter $\alpha$, but both peak values of $u$ and $T_{e}$ sharply
     increases with the parameter $\alpha$. These result in that the presence of the parameter $\alpha$ enhances the higher radiative flux in the images and they are also shown in Fig.\ref{fig:4} where we present the radiative flux of the simulated images in Fig. \ref{fig:3}.
    Therefore, in the black hole images, the MOG parameter $\alpha$ decrease the equilibrium tori size from the spacetime effects \cite{2023ApJ...942...47Y},  and simultaneously increases the total flux from plasma effects \cite{2022ApJ...941...88O}.
     \begin{figure}
        \centering
        \includegraphics[width=16cm]{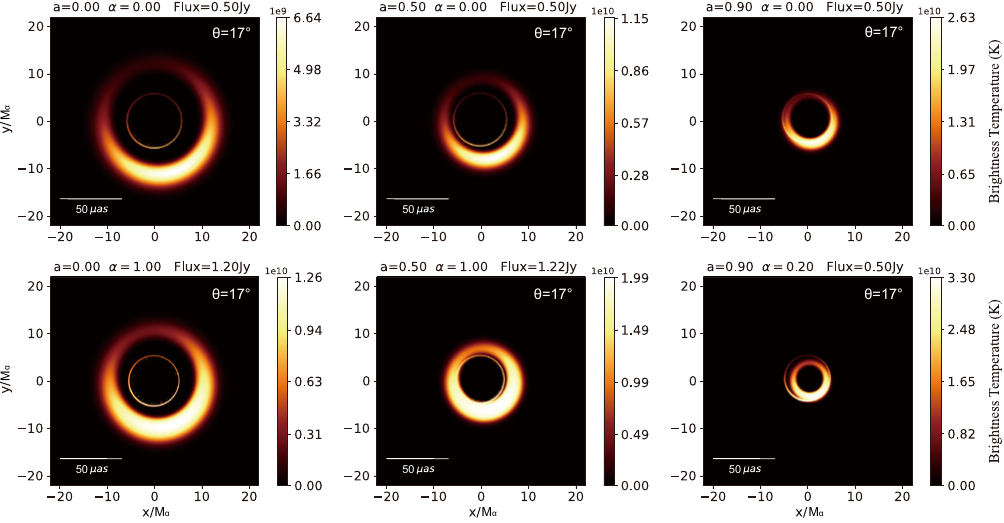}
        \caption{Simulated images of geometrically thick magnetized equilibrium tori around a Kerr-MOG black hole with different MOG parameter $\alpha$ and spin parameter $a$. The observed inclination is set to $\theta=17^\circ$ and the position angle (PA) is $288^\circ$. The rescale parameter $\mathcal{M}$ for different spin parameters are adjusted to ensure that images with $\alpha=0$
        have the identical radiative flux value $0.5 \ \mathrm{Jy}$.} 
        \label{fig:5}
    \end{figure}

   In order to compare with the observed image of the black hole M87*, in Fig. \ref{fig:5}, we presents simulated images of geometrically thick magnetized equilibrium tori around a Kerr-MOG black hole with an observation inclination angle $\theta=17^\circ$ and a position angle (PA) of $288^\circ$ \cite{2018ApJ...855..128W,2023Natur.621..711C}.Here, the rescale parameter $\mathcal{M}$ for different spin parameters are adjusted to ensure that images with $\alpha=0$ have the identical radiative flux value $0.5 \ \mathrm{Jy}$ .
   In this special case with observation inclination angle $\theta=17^\circ$, we find that the changes of equilibrium tori images with the MOG parameter are similar to those shown in Fig. \ref{fig:3}. However, we also find that in the cases with non-zero $\alpha$, the image's size decays more rapidly with the spin parameter $a$.
   \begin{figure}
        \centering
        \includegraphics[width=16cm]{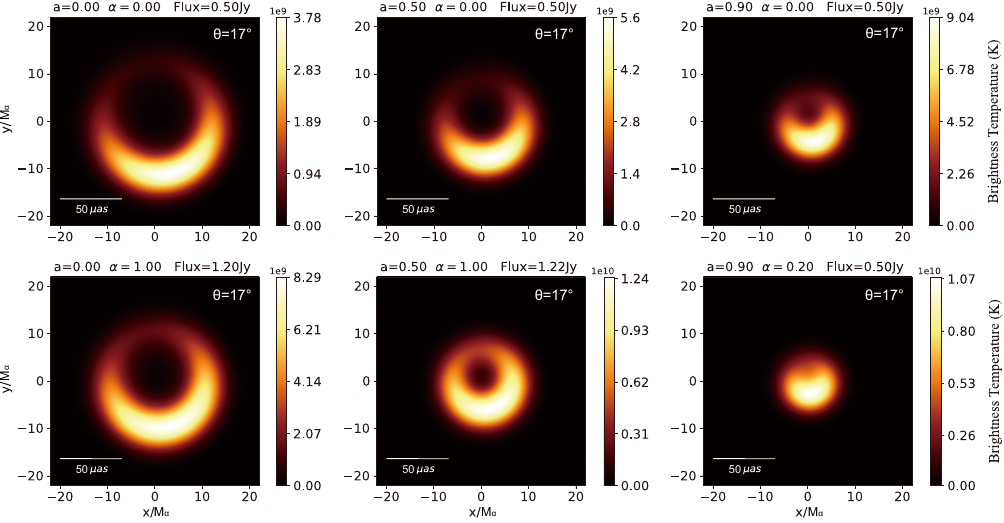}
        \caption{Blurred images of geometrically thick magnetized equilibrium tori around a Kerr-MOG black hole for the images in Fig. \ref{fig:5} through convolving with a $20\mu as$ FWHM Gaussian beam.}
        \label{fig:6}
    \end{figure}

\begin{figure}
        \centering
        \includegraphics[width=16cm]{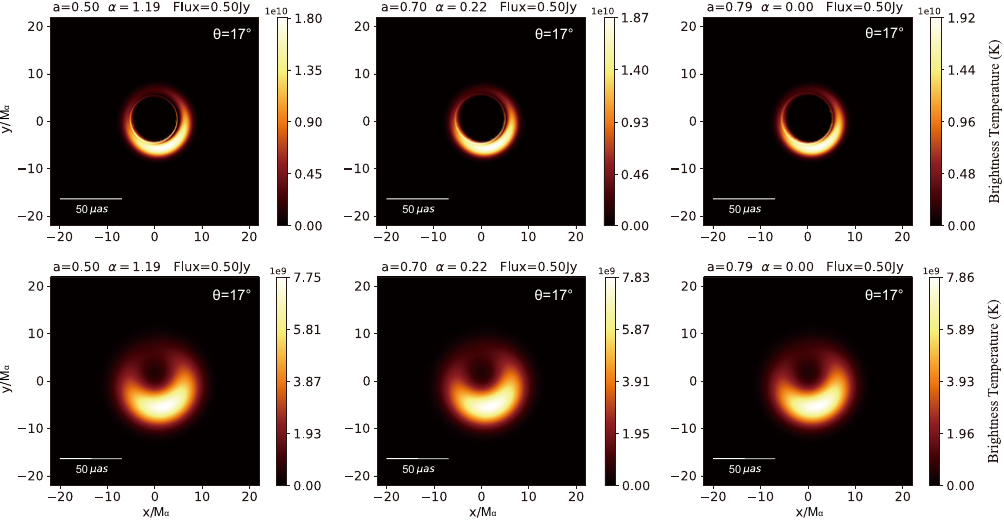}
        \caption{Highly overlapping of Kerr-MOG black hole images for three distinct parameter sets of $a$ and $\alpha$. Here the diameters of the bright rings
        are set $d=43.3\mu as$. The rescale factor $\mathcal{M}$ was adjusted to that the flux is identical to  $0.5 \  \mathrm{Jy}$ for each parameter sets.}
        \label{fig:7}
    \end{figure}

   In Fig. \ref{fig:6}, we also presents blurred images of geometrically thick magnetized equilibrium tori around a Kerr-MOG black hole for the images in Fig. \ref{fig:5} through convolving with a $20\mu as$ FWHM Gaussian beam, which is convenient to compare with the current EHT observational images for the black hole M87* due to its limited resolution \cite{2019ApJ...875L...1E,2019ApJ...875L...5E,2024A&A...681A..79E}.
   As illustrated in Fig. \ref{fig:6}, the thin bright ring in the image disappears due to the blurring process, and the crescent-shaped bright zone becomes more broaden, which yields that features of the blurred images are highly consistent with those of observed black hole images \cite{2019ApJ...875L...1E,2024A&A...681A..79E}.
   The size of the bright rings is evidently influenced by the black hole spin $a$ and the MOG parameter $\alpha$, which implies that images of Kerr-MOG black hole can be compatible with the observed image of the black hole M87* by tuning these two parameters. However, we find that images of Kerr-MOG black hole are highly overlapping for certain parameter sets of $a$ and $\alpha$, so that the simulated images  are practically indistinguishable from each other,
   which is shown in Fig. \ref{fig:7}. Here, the rescale factor $\mathcal{M}$ was adjusted to that the flux is identical to  $0.5 \  \mathrm{Jy}$ for each parameter sets.
 Moreover, for these highly overlapping Kerr-MOG black hole images for three different parameter sets, the corresponding equilibrium tori share the identical electron density $n_e \sim 2 \times 10^6 \  \mathrm{cm}^{-3}$, electron temperatures $T_e \sim 3 \times 10^{10} \  \mathrm{K}$, and magnetic field strengths $B \sim 11 \  \mathrm{G}$.  Especially, our results also show that  the Kerr black hole images can overlap with those of Kerr-MOG black holes with the lower spin parameter $a$ due to the presence of the MOG parameter $\alpha$. These overlapping images  lead to a challenge for us to identify black hole parameters by using of black hole images.
    \begin{figure}
        \centering
        \includegraphics[width=16cm]{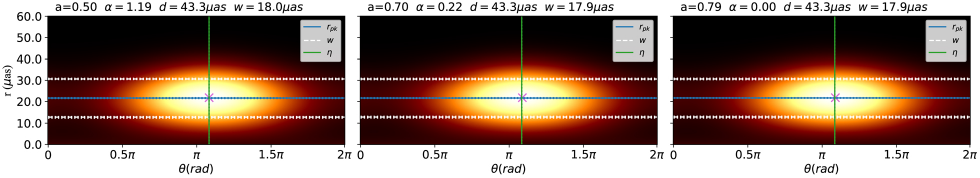}
        \caption{Unwrapped ring profiles of the simulated images in Fig. \ref{fig:7}. The estimated radius $d/2$ is shown with a horizontal line, with dotted lines denoting the associated uncertainty. Horizontal dashed lines at $(d\pm w)/2$ show the measured ring width. Vertical green lines give the orientation angle $\eta$  and its uncertainty. The purple cross marks the peak brightness in each panels.}
        \label{fig:8}
    \end{figure}
    Following the methodology in \cite{2019ApJ...875L...4E}, in Fig. \ref{fig:8}, we present the unwrapped ring profiles of the simulated images, which clearly show that Kerr-MOG black hole images for three different parameter sets selected in Fig. \ref{fig:7} have the same diameters of the bright rings $d=43.3 \ \mu as$ as observed in the black hole M87* images \cite{2024A&A...681A..79E}. As in ref \cite{2019ApJ...875L...1E, 2019ApJ...875L...4E}, the ring widths in the simulated images are approximately $w\sim 18\ \mu \mathrm{as}$, which falls within the observational constraint $w< 20 \ \mu as$.
    \begin{figure}
        \centering
        \includegraphics[width=8cm]{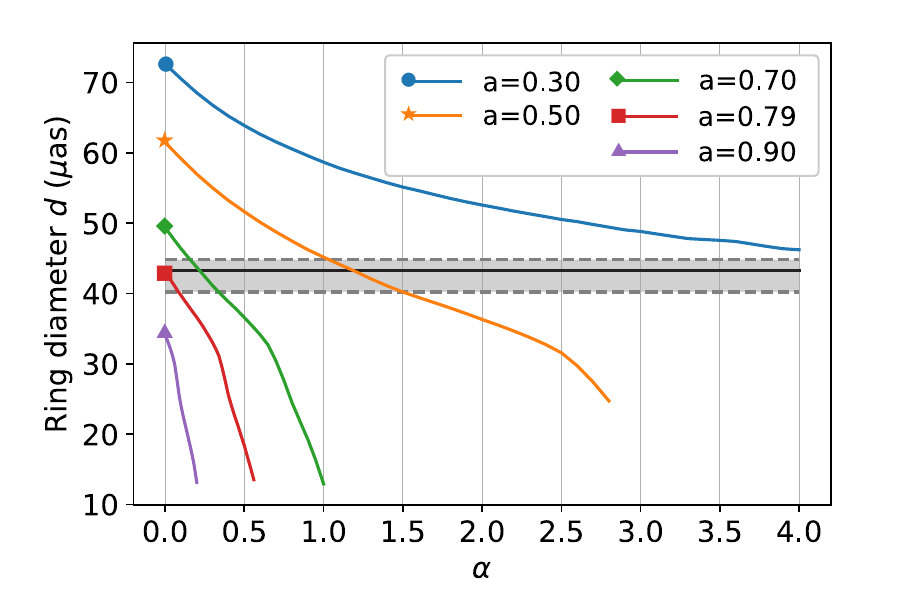}
        \caption{Ring diameters varies with the $\alpha$ parameter for different spin parameters $a=0.3$, $0.5$, $0.7$, $0.79$ and $0.9$. The observed $43.3 \ \mu \mathrm{as}$ diameter in M87* \cite{2024A&A...681A..79E} is denoted by a central black line and the gray region corresponds to the 1-sigma uncertainty range.}
        \label{fig:9}
    \end{figure}
    Fig. \ref{fig:9} presents the allowed range of the MOG parameter $\alpha$ and spin parameter $a$ for the Kerr-MOG black hole in term of the bright ring diameter of the M87* images. Here, the central black line represents the M87* observed ring diameter as $43.3 \ \mu \rm{as}$ and the gray region denotes the 1-sigma uncertainty range
    \cite{2024A&A...681A..79E}. Fig. \ref{fig:9} shows that the presence of the MOG parameter $\alpha$ broadens the allowed range of the spin parameter $a$. Moreover, we also note that the very high black hole spin cases are not favorite in this work, which is inconsistent with those of EHT. This difference is mainly caused by  that here the accretion disc is a kind of geometrically thick magnetized equilibrium tori, which exists certain deviations from the real accretion disc around black holes in our universe.
    \begin{figure}
        \centering
        \includegraphics[width=16cm]{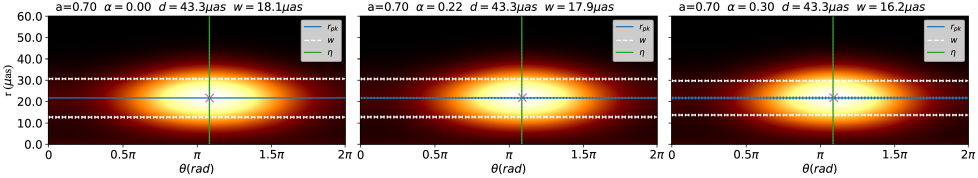}
        \caption{Unwrapped ring profiles of the simulated images for black hole masses of \(5.78 \times 10^9 M_{\odot}\), \(6.5 \times 10^9 M_{\odot}\), and \(6.78 \times 10^9 M_{\odot}\), and corresponding \(\alpha\) values of $0$, $0.22$, and $0.3$ (from left to right), with \(a = 0.7\). The rescale factor \(\mathcal{M}\) was adjusted to maintain an identical flux  \(0.5 \ \mathrm{Jy}\) in each panel.}
        \label{fig:10}
    \end{figure}

Fitting with the observed image, the estimated black hole mass is found to be much smaller than the value measured by EHT if the emission is produced at comparatively large radius in the disk \cite{2019ApJ...875L...5E}. From Fig. \ref{fig:10}, with the observed bright ring $d = 43.3 \mu \rm{as}$ in the black hole M87* image, we find that the estimated black hole mass is $5.78 \times 10^9 M_{\odot}$ in the case with the fixed $a=0.7$ and $\alpha=0$, which is  much smaller than the value $6.5 \times 10^9 M_{\odot}$ measured by EHT. However, in the case with $a=0.7$ and $\alpha=0.22$, the estimated black hole mass in our fitting becomes $6.5 \times 10^9 M_{\odot}$, which is identical to that measured by EHT. Furthermore, for the fixed $a=0.7$ and $\alpha=0.3$, the corresponding estimated black hole mass becomes $6.78 \times 10^9 M_{\odot}$, which is larger than that measured by EHT. The main reason is that the increasing value of $\alpha$ decreases the size of the equilibrium tori, which yields that the emission in the spacetime shifts gradually from the comparatively far region to that near the black hole.
% \underline{so the emission produced region gradually changes from the comparatively large radius to the small one}.

    \section{Summary}
    \label{sec:4}

    We have firstly studied geometrically thick magnetized equilibrium tori around a Kerr-MOG black hole with an extra MOG parameter $\alpha$. We find that both the black hole spin $a$ and the MOG parameter $\alpha$ lead to the sharp decrease in the size of equilibrium tori, but the shape of equilibrium tori is not susceptible to these two parameters. Moreover, the rest mass density distribution is similar to that of the potential $w$, but the rest mass density $\rho$ own more rapidly rate of decay with the distance from the center. Then, we simulated the images of Kerr-MOG black holes surrounded by the geometrically thick magnetized equilibrium tori. The images show that there are closely-packed bright thin rings together with a crescent-shaped bright zone. With the increasing of the MOG parameter $\alpha$, both the brightness temperatures and the radiative flux increase, but the size of the images decreases. This behavior contrasts with that of geometrically thin accretion disks \cite{2017PhRvD..95j4047P,2023EPJC...83..264H}, primarily due to differences in the accretion disk models and radiation mechanisms. Moreover, we also find that the width of the crescent-shaped bright zone increases with $\alpha$. Due to the presence of the MOG parameter $\alpha$,  the image's size decays more rapidly with the spin parameter $a$. Especially, we find that images of Kerr-MOG black hole can be highly overlapping for certain parameter sets of $a$ and $\alpha$, which lead to a challenge for us to identify black hole parameters by using of black hole images. However, differences in the power of relativistic jets between Kerr-MOG black holes and standard Kerr black holes may potentially break this indistinguishability \cite{2024JCAP...01..047C}. Further, we presents the allowed range of the MOG parameter $\alpha$ and spin parameter $a$ for the Kerr-MOG black hole in term of the bright ring diameter of the M87* images, which shows that
    the presence of the MOG parameter $\alpha$ broadens the allowed range of the spin parameter $a$.  These results could help us to deeply understand black hole images and the Kerr-MOG black hole in the scalar-tensor-vector gravity theory.

    \section{\bf Acknowledgments}

    This work was supported by the National Natural Science Foundation of China under Grant No.12275078, 11875026, 12035005, 2020YFC2201400, and the innovative research group of Hunan Province under Grant No. 2024JJ1006.

    \bibliography{kerrmog}

\end{document}